\documentclass[anonymous]{IEEEtran}
\IEEEoverridecommandlockouts
\usepackage[utf8]{inputenc} 
\usepackage[T1]{fontenc}    
\usepackage{hyperref}       
\usepackage{url}            
\usepackage{booktabs}       
\usepackage{amsfonts}       
\usepackage{nicefrac}       
\usepackage{microtype}      
\usepackage{graphicx}			
\usepackage{epsfig}
\usepackage{floatrow}
\usepackage{tabu}
\usepackage{cite}
\usepackage{amsmath,amssymb,amsfonts}
\usepackage{algorithmic}
\usepackage{textcomp}
\usepackage{xcolor}
\usepackage{amsmath}
\usepackage[ruled]{algorithm2e}
\algsetup{linenosize=\small}
\def\BibTeX{{\rm B\kern-.05em{\sc i\kern-.025em b}\kern-.08em
    T\kern-.1667em\lower.7ex\hbox{E}\kern-.125emX}}

\usepackage{graphics} 
\usepackage{times} 
\usepackage{color}
\usepackage{multirow}

\usepackage{xspace}
\makeatletter
\DeclareRobustCommand\onedot{\futurelet\@let@token\@onedot}
\def\@onedot{\ifx\@let@token.\else.\null\fi\xspace}
\let\labelindent\relax

\newcommand{\Cref}[1]{Chap.~\ref{#1}}

\title{Learning Personalized Page Content Ranking Using Customer Representation}

\author{%
 Xin Shen, Yan Zhao, Sujan Perera, Yujia Liu, Jinyun Yan, Mitchell Goodman}

\begin{document}

\maketitle

\thispagestyle{plain}
\pagestyle{plain}

\begin{abstract}
On E-commerce stores, there are rich recommendation content to help shoppers shopping more efficiently.  However given numerous products, it's crucial to select most relevant content to reduce the burden of information overload. We introduced a content ranking service powered by a linear causal bandit algorithm to rank and select content for each shopper under each context. The algorithm mainly leverages aggregated customer behavior features, and ignores single shopper level past activities. We study the problem of inferring shoppers interest from historical activities. We propose a deep learning based bandit algorithm that incorporates historical shopping behavior, customer latent shopping goals, and the correlation between customers and content categories. This model produces more personalized content ranking measured by 12.08\% nDCG lift. 
\end{abstract}

\begin{IEEEkeywords}
Information systems; Content ranking; Personalization; Applied computing; Online Shopping; Content diversity; Causal bandit; Contextual bandit; View-through attribution; Holistic optimization
\end{IEEEkeywords}
\section{Introduction}
E-commerce stores like Walmart and eBay generate product recommendations to facilitate the shopping experience. These recommendations are often rendered in a carousal widget. For example, lighting deals widget collects transient deals for shoppers to explore; frequently bought together widget shows what other products shoppers often purchase together for shoppers to consider to buy it together; compare with similar widget lists alternative products with fine grained comparison for shoppers to make decision. Such recommendations distinguish online shopping from offline store shopping, where shoppers have to research related content manually. However too much recommendation content can be overwhelming and irrelevant content can be trust-bust. To select the most relevant content, we build a content optimization system to rank widget candidates and determines which content to display when a customer visits our online store. This system balances differing business objectives and methods through optimization capabilities and considers customer, content, and shopping context. 

We observed customers show unique preference on content. For example, for deal-hunter and price sensitive customers, they would like to examine products in the deals content; for customers have preferred brand, they engage more with contents containing products from the same brand or similar brands. The current ranking algorithm includes customers profile features and aggregated behavior features, ignoring rich in-session activities. These activities (e.g. view, click, purchase, comment on products) implicitly carry shopping interest and intent. In this paper, we study the problem to improve ranking efficiency by learning shopping interest  from in-session activities.

To infer in-session shoppers interest, it requires a dynamically learnable process to correlate the customers' engagement, customers' representation with the different contents. The challenges are two-fold:
\begin{itemize}
\item For correlation operations to be performed, the customer representation and content representation must be aligned within the same embedding space. Concatenating or pooling feature embeddings to aggregate latent engaged product representation may be naive and lack guarantees. Therefore, the question is how to obtain these representations while thoroughly considering customers' shopping patterns and the main characteristics of the content that reveal customers' preferences.
\item To optimize our problem, it is crucial to learn the latent correlation between customer representation, content representation, and other categorical content contexts, such as brand. Traditional approach \cite{hirata2022solving} that uses simple inner-products to determine correlation conditions may be biased since the implicit reasoning behind why recommended content captures a customer's interest is not deterministic. Therefore, a learnable process is necessary for this optimization work.
\end{itemize}

\textbf{Our contributions} Our work introduces a neural bandit approach for content optimization that prioritizes ranked content based on customers' shopping preferences and missions while contributing to the overall commercial optimization goal. The technical contributions are as follows:
\begin{enumerate}
    \item We proposed solution to use customers' shopping pattern-aware product category presentation\cite{hao2020p} and granular item embedding\cite{muhamed2021ctr} to produce content representation and customer representation. This ensures that both representations are in the same semantic embedding space and contain high-level customers' shopping intent and granular shopping preferences.
    
    \item Our new application idea involves using a channel-wise attention-based convolutional neural network for cross-feature optimization work. This approach dynamically learns correlations among different feature representations while optimizing the commercial goal.
\end{enumerate}


\section{Related Works}
\subsection{Causal Bandit to Content Optimization}
Application of exploration strategies in the context of recommender systems is an active area of research. In recent years, both linear and deep neural network approaches have emerged in recent years and shown promising results~\cite{bietti2021contextual,riquelme2018deep} applied by e-commerce. 
To optimize business value, it is known to use a causal bandit optimization approach we developed \cite{kanase2022An} which considers content type, customer type, etc. In this work, the focus is on achieving personalized content ranking performance by leveraging customer embeddings, content representations, and customer-to-content correlations for a more effective bandit optimization approach with the goal of helping customers fulfill their shopping goals to further optimize business impact. 

\subsection{Customer Representation and Product Embedding}
Efficient encoding of customer shopping patterns and engagements into a dense representation aims to capture each customer's implicit shopping intent in a latent space. These representations have proven useful for personalization and recommendation~\cite{ji2003online,goossens2018effective,abdul2021commerce,park2009individual}. Most of the existing methods~\cite{xu2022rethinking, pal2020pinnersage} have introduced an efficient approach to aggregate customer's representation while ensuring such representation is within the same products' embedding space to perform tasks, such as products' recommendation retrieval and similar customer's retrieval. Our work largely refer to these approaches to build our customer representation and content embedding based on our latest work to build shopping-intent aware product embeddings. 

\subsection{Multi-Channel Deep Neural Network}

Convolutional neural networks (CNNs) have made significant progress in recent years~\cite{goodfellow2016convolutional, goodfellow2016deep}. Research has demonstrated that CNNs can encode a broad range of feature-maps, not just limited to images, into a dense and intricate representation in a higher manifold~\cite{gu2018recent}. This rich representation is beneficial for many downstream optimization tasks, such as recognition, regression, and question answering. Considerable research efforts have been directed towards improving the recognition performance of Convolutional Neural Networks (CNNs), as evidenced by the extensive studies~\cite{shen2023data, xin2020complex, al2017review}. However, there has been a noticeable dearth of emphasis on harnessing the potential of CNNs for regression tasks. Recent works have shown that increasing the feature representation via a multi-channel approach can significantly improve a model's generalization abilities for regression tasks~\cite{shen2023fishrecgan, shen2023semantic}, but with its own limit. To learn such a representation more robustly, the development of channel attention mechanisms has contributed to the effectiveness of CNNs in dynamically learning the correlations between different layers of feature maps~\cite{zhang2022resnest, zhang2018image, chen2017sca}. Channel attention mechanisms are particularly useful for optimizing tasks with a multi-channel input feature map. We introduce a novel application of the channel-wise attention network to perform multi-channel-feature-based optimization for our content ranking job. Our experimental results show that our approach outperforms existing methods and demonstrates the effectiveness of the channel-wise attention network in content ranking tasks. 


\section{Problem Formulation}

Today's recommender system collects the interactions between each user $\mathcal{U}$ and item $i$ , and sorts the items based on the relevance to each customer's shopping behaviors (i.e., clicks, views, preferences). Similarly, for our content optimization use case, we can view the entire process as a broader recommender system model, and the recommened items are the store content widget groups $\mathcal{B}$. Model needs to correlate each customer to different contents for personalized ranking purposes. We assume by providing a more personalized and relevant content to each customer would contribute to our business goal $\mathcal{R}$. The optimization objective of this problem can be formalized as:

\begin{equation}
    \mathbb {E}(\mathcal{R} | \mathcal{U}, \mathcal{B}),
\end{equation}
which optimizes the entire store's business goal $\mathcal{R}$ under the condition of ranking the personalized content widget groups for each individual customer.

\section{Methodology}

In this section, we introduce a deep learning-based bandit framework for content optimizatio. Our approach involves building personalized features and modeling multiple features as multi-bandits to rank contents while optimizing our business goal \textit{MOI}. 

\subsection{Features}
To perform our content optimization work while enhancing the personalization capacity, we introduce 4 types of features of [$\beta$, $\lambda$, $\tau$, $\gamma$] contributing to categorical bandit and latent personalized bandit respectively. Please the detailed explanation regarding these features in the corresponding sub-sections below.

\subsubsection{Static Categorical Bandit}
As in \cite{kanase2022An}, when a customer visits web page $p \in \mathcal{P}$ on store’s retail website, we receive customer context $\mathcal{C}$ and shopping context $\mathcal{X}$. Context $\mathcal{Z}$ corresponding to each candidate content can also be generated separately. We
then combine contexts $\mathcal{C}$, $\mathcal{X}$ and $\mathcal{Z}$ non-linearly to form a single d-dimensional vector $\beta$ $\in \mathcal{R}^{d}$. We call this bandit $\beta$ as the categorical feature. For reference, we include a few examples of context below:

\begin{itemize}
  \item Shopping context: region, web page type, widget group id,
page item, metadata of page item, and search query
  \item Customer context: recent interaction events, customer signed-in status, and prime membership status
  \item Content context: widget id, widget meta information, and
content attributes
\end{itemize}

\subsubsection{Latent Personalized Bandit}



To obtain two customer representations $\lambda$ and $\tau$, following the similar setup \cite{pal2020pinnersage}, we aggregate two recently developed product embedding technologies on the product category level \cite{hao2020p} and item level \cite{muhamed2021ctr} respectively. These embeddings capture product semantics and shopper behavior, such as co-viewed and co-purchased items. The customer representation $\lambda$ from the product category level aggregation, provides a high-level signal of a customer's overarching shopping goal, while $\tau$, the item level aggregation, captures granular preferences such as color and brand.


For content representation $\gamma$, we aggregate all products in a content widget to ensure $\gamma$ is within the same embedding space as the customer representations. For example, the "Buy Again" widget contains four products, and we use the product category level embedding to encode the widget's overall characteristics and purpose.

In the modeling step, all [$\lambda$, $\tau$, $\gamma$] serve as the latent personalized bandit features.

\subsection{Reward}
We propose optimizing for overall down-session value generated by customer interaction with content, consistent with the existed solution\cite{kanase2022An}. Once ranked and rendered, we record the customer's interaction events such as impressions, clicks, purchases, and other high-value actions. The aggregate value generated over a subsequent time horizon is then used to compute our metric of interest, \textit{MOI}. Notably, such MOI can be fetched in real-time. Content meeting a predefined criteria is attributed with this measured value as reward. Content ranking models then learn to predict this down-session value when showing content to customers in a given context and make ranking decisions accordingly. Our approach measures and attributes site-wide impact across all devices, apps, widget groups, and web pages from the moment of customer interaction with content, and we refer to it as a holistic optimization of diverse content types using aggregate down-session value.

\subsection{Ranker Model}

Our objective is to rank each eligible candidate c in a set of candidate contents $C_q$, such that the \textit{top}- K ranked contents $T_q$ are personalized according to each customer's shopping preference and intent, while optimizing the desired reward $\mathcal{R}$. Our proposed approach, differs from the linear modeling approach described in our previous work\cite{kanase2022An}, which models the correlation between $\mathcal{R}$ and categorical bandit features via a generalized linear model. Instead, we propose a channel-wise-attention deep neural network approach, ResNeSt\cite{zhang2022resnest}, to model the mapping of both categorical bandit and latent personalized bandit to the $\mathcal{R}$. We allow this model's cross-channel attention mechanism to dynamically learns the latent correlation among all features in different channels, especially the semantic correlation among content representation $\gamma$, customer presentation $\lambda$ and $\tau$, to fuse all bandits into a dense representation in the higher-manifold. The dense representation is then used to optimize the designated $\mathcal{R}$. Compared to many existing content retrieval and searching approaches ~\cite{pal2020pinnersage, hirata2022solving}, which use the inner-product to calculate the ranking score between the customer representation and the product embedding, our approach learns such correlation dynamically while optimizing upon the $\mathcal{R}$, which represents the desired business metric MOI.

\section{Model evaluation and experiments}


In this section, we evaluate our proposed approach to study the following research questions (RQs) on our proposed novelties: 

\noindent $\bullet$\textbf{RQ1:} Do the proposed personalization features, including customer representations and content representations ($\mathbb{\lambda}$, $\mathbb{\tau}$, and $\mathbb{\gamma}$), improve the personalized ranking performance? 

\noindent $\bullet$\textbf{RQ2:} Is the split-channel-wise-attention mechanism essential for a deep neural bandit model to utilize personalization features?

\subsection{Offline results} 
\subsubsection{Evaluation Protocol}
We evaluate all models with the following metrics:
\begin{enumerate}
  \item \textbf{Regression metrics}. 
    Our problem is a regression problem where we regress on observed reward $\mathcal{R}$ and then rank contents. Although lower regression errors don't always guarantee better ranking, models should aim for lower regression errors to improve ranking while ensuring interpretability. We propose two widely used regression metrics: (1) \textbf{MSE Loss}, which measures the quality of regression with a bias-variance trade-off and is also our model's training loss function; (2) \textbf{MAE}, which measures the quality of regression similar to MSE but is less sensitive to outliers that often occur in production systems. Both further evaluate how model is fitted to the data to justify if there exists correlation between features and $\mathcal{R}$.
  \item \textbf{Ranking metrics}. The ultimate goal of our model is to rank contents. In this work, we propose to use \textbf{NDCG@5,} A position-aware metric measuring quality of final ranking, which assigns larger weights for higher positions; The page we are evaluating against only has 5 slots, so we limit our NDCG measurement to top 5. In this work, to evaluate against improvements on personalized ranking, we define engagement reward $Re$ as customers click on a content and also generate down-session reward $R$ (our metric of interest), so $Re = I(clicked) * R$, NDCG@5 is thus defined as $\frac{DCG_5}{IDCG_5}$, with $IDCG_5$ being ideal order of $DCG_5$ and
  
  \begin{equation}
    DCG_5 = \sum_{i=1}^{5} \frac{Re_i}{log_2(i + 1)},
\end{equation}

\end{enumerate}

We report the averaged metrics over all users.
We select the model to report the test set performances based on the best validation NDCG@5 score.

\subsubsection{Baselines}
\begin{table*}[t]
  \caption{Table for offline results}
  \label{table:result}
  \resizebox{\textwidth}{!}
  {
  \label{offline-results}
  \centering
  \begin{tabular}{lllll}
    \toprule
    \multicolumn{5}{c}{\textit{All Compared to Production Baseline : A Generalized Linear Model Using Static Categorical Features}}\\
    \midrule
    \midrule
    Label   & Methods   & MSE Loss Reduction  & MAE Reduction  & nDCG@5 Improvement \\
    \midrule

  \multicolumn{5}{l}{\textit{Using Static Categorical Features for Customers and Contents}} \\
  \midrule
  1. NN baseline A & Feed-forward Neural Network model with categorical features &  
  +2.78\% & -4.25\%  & +5.59\%  \\   
  
  2. DIN-A  & DIN with categorical features	 &  
  -3.78\% & +4.27\%  & -16.08\%  \\    
  \midrule
  \multicolumn{5}{l}{\textit{Using Both Static Categorical Features and Personalized Features for Customers and Contents}}\\
  \midrule
  3. NN baseline B  & NN baseline with our personalization features and categorical features	 &  
  -6.70\% & -1.68\%  & -0.99\%  \\
  4. DIN-B  & DIN-A with with our personalization features and categorical features	 &  
  -26.58\% & -27.59\%  & -34.27\%  \\    
  \textbf{5. Our work}  & \textbf{ResNeSt based Neural Bandit with our personalization features and categorical features}	  &  
  \textbf{-19.61\%} & \textbf{-8.15\%}  & \textbf{+16.08\%}  \\   

    \bottomrule
  \end{tabular}
  }
\end{table*}
We compare our proposed model with baselines and conduct experiments to see the impact of (1) introducing the personalization features, and (2) how different model structure would be adaptable to personalization features along with the existed categorical features:

\begin{enumerate}
    \item \textbf{Production baseline}:
    This approach is a static recommendation method, currently in production. We used Bayesian Linear Bandit that uses static categorical features, including customer context, such as "is\_prime\_member", and static content features to rank content with an exploration/exploitation trade-off.
    
    \item \textbf{NN baseline A}:
    As the recent work in neural bandit \cite{xu2020neural} indicates its robustness in deep representation for exploration, we implemented a feed-forward Neural Bandit approach without attention mechanism \cite{vaswani2017attention} using same features as production baseline does. This is to evaluate differences between linear and non-linear approach under our problem setup
    
    \item \textbf{DIN-A}:
    We further leveraged a state-of-the-art (SOTA) attention-recommendation model Amazon recently introduced~\cite{muhamed2021ctr}, which is able to capture the interactions among the customers' engagements. The input features stay the same as the production baseline's. We want to evaluate the performance lift via a more complex model structure with attention mechanism.

    \item\textbf{NN baseline B}
    This is the same model structure as \textit{NN baseline A} using both production's categorical feature and the newly proposed personalized feature We aim to evaluate incremental benefit of including more personalized features in a simple solution

    \item\textbf{DIN-B} We utilized the same SOTA attention-model structure with both production's categorical features AND our propose personalization features. This is to evaluate incremental benefit of including more personalized features in the SOTA model structure
    
    \item\textbf{ResNeSt Neural Bandit}: This is proposed model structure with the desired split-channel-attention mechanism~\cite{zhang2022resnest} to properly learn the interactions between both categorical features and personalization features for our content optimization work
\end{enumerate}

\subsection{Performance Comparison}

Table~\ref{table:result} shows our proposed solution and baseline performance across all benchmarks. Using ResNeSt architecture with static categorical and personalized features, our solution consistently performed well across all metrics, with significant MSE loss (-19.61\%) and MAE (-8.15\%) reductions. Our solution also significantly improved nDCG compared to customers' real data on preferred content, demonstrating feasibility for high-relevant personalized content ranking while optimizing the desired reward $\textit{R}$. Analysis below justifies the necessity of both proposed contributions in personalized features and the application of ResNeSt as the model architecture.

\noindent \textbf{Personalized Feature Importance, RQ1}. 
Experiment 1 and 2 show consuming only static categorical features with robust model architecture fails to improve performance. \textit{NN baseline A} in Experiment 1 improved nDCG (+5.59\%) but inconsistent MSE and MAE reductions suggest failure to fit categorical features. Weak correlation between features and optimization objective $\mathcal{R}$ leads to loss of fidelity in ranking performance. Experiment 2 with \textit{DIN-A} using SOTA model structure showed similar results, with worse personalized performance (-16.08\% drop in nDCG).

On the other hand, Experiment 3 (NN baseline B) and Experiment 4 (DIN-B) showed consistent improvements in model fitting performance when fed with our proposed personalization features, with decreased MSE loss and MAE. This manifest that our proposed personalization features shows a stronger correlation towards the end optimization reward $\mathcal{R}$

\noindent \textbf{Model Structure Importance, RQ2}. 
Experiments 3 (NN baseline B) and 4 (DIN-B) learned via stronger correlation between personalized feature and reward R, but failed in personalized ranking (-0.99\% and -34.27\% in nDCG). This shows that simple deep neural net and DIN with traditional attention layers didn't learn personalized features properly. Notably, the proposed ResNeSt model with split-channel-wise attention enabled robust performance in model fitting and personalized ranking, proving the necessity of our proposed model structure for proper personalized feature learning.



\section{Conclusion and Future Work}
Today's E-commerce's shopping pages contain various content types serving different shopper intents. To rank these content types effectively, it is essential to understand individual shopper behavior, which the current content ranking solution does not consider. To address this, we propose a methodology that utilizes both coarse and fine-grained product embeddings to represent shopper preferences, resulting in improved content ranking performance compared to state-of-the-art methods. Our results demonstrate the potential for improving e-commerce content ranking by understanding shopper intent through their interactions. In the future, we plan to enhance product embeddings by incorporating richer information, such as users' interaction knowledge graph as a prior and richer catalog data, to develop a methodology that fuses the coarse and fine-grained embeddings for content ranking~\cite{chen2022attentive, lkgr, li2023text}.
\newpage
\bibliographystyle{plain}
\bibliography{reference}

\clearpage
\end{document}